\documentclass[apjl]{emulateapj}
\usepackage{natbib}
\usepackage{amsmath}

\shorttitle{Combining Spitzer parallax and Keck II adaptive optics
}
\shortauthors{Beaulieu et al.}

\begin{document}


\title{
        Combining Spitzer parallax and Keck II adaptive optics imaging to measure the mass
        of a solar-like star orbited by a cold gaseous planet discovered by microlensing
         }


\author{J.-P.~Beaulieu\altaffilmark{1,2}, 
 V.~Batista\altaffilmark{2}, D.P.~Bennett\altaffilmark{3},  J.-B.~Marquette\altaffilmark{2,4}, J.W. Blackman\altaffilmark{1}, A.A.Cole\altaffilmark{1}, 
 C. Coutures\altaffilmark{2},
C. Danielski\altaffilmark{5}, D. Dominis-Prester\altaffilmark{6}, J. Donatowicz\altaffilmark{7}, A.~Fukui\altaffilmark{8},  N. Koshimoto\altaffilmark{9}, C. Lon\v{c}aric\altaffilmark{6},  J.C. Morales\altaffilmark{10},  
 T.~Sumi\altaffilmark{7},  D.~Suzuki\altaffilmark{3},  C. Henderson\altaffilmark{11,12,13}, Y. Shvartzvald\altaffilmark{11,14} and C. Beichman\altaffilmark{12}
}

\altaffiltext{1}{School of Physical Sciences, University of Tasmania, Private Bag 37 Hobart, Tasmania 7001 Australia; 
Jeanphilippe.beaulieu@utas.edu.au;Andrew.Cole@utas.edu.au}
\altaffiltext{2}{Sorbonne Universit\'es, UPMC Universit\'e Paris 6 et CNRS, UMR 7095, Institut dÕAstrophysique de Paris, 98 bis bd Arago, 75014 Paris, France; beaulieu@iap.fr}
\altaffiltext{3}{Laboratory for Exoplanets and Stellar Astrophysics, NASA/Goddard Space Flight Center, Greenbelt, MD 20771, USA}
\altaffiltext{4}{Laboratoire d'Astrophysique de Bordeaux, Univ. Bordeaux, CNRS, B18N, all\'ee Geoffroy Saint-Hilaire, 33615 Pessac, France}
\altaffiltext{5}{GEPI Observatoire de Paris, Section de Meudon 5, place Jules Janssen 92195 Meudon, France}
\altaffiltext{6}{Department of Physics, University of Rijeka, Radmile Matej\v{c}i\' c 2, 51000 Rijeka, Croatia}
\altaffiltext{7}{Technische UniversitŠt Wien, Karlsplatz 13, 1040 Wien, Austria}
\altaffiltext{8}{Okayama Astrophysical Observatory, National Astronomical Observatory of Japan, Asakuchi, Okayama 719-0232, Japan}
\altaffiltext{9}{Department of Earth and Space Science, Graduate School of Science, Osaka University, 1-1 Machikaneyama, Toyonaka, Osaka 560-0043, Japan}
\altaffiltext{10}{Institut de Cincies de lÕEspai (CSIC-IEEC), Campus UAB, Carrer de Can Magrans s/n, E-08193 Cerdanyola del Valls, Spain;morales@ieec.uab.es}
\altaffiltext{11}{Jet Propulsion Laboratory, California Institute of Technology, 4800 Oak Grove Drive, Pasadena, CA 91109, USA}
\altaffiltext{12}{NASA Exoplanet Science Institute, California Institute of Technology, Jet Propulsion Laboratory, Pasadena, CA 91125, USA}
\altaffiltext{13}{PAC/NExScI, Mail Code 100-22,  Caltech, 1200 East California Boulevard, Pasadena, CA 91125}
\altaffiltext{14}{NASA Postdoctoral Program Fellow}
\begin{abstract}
To obtain accurate mass measurements for cold planets discovered by microlensing, it is usually necessary to 
combine light curve modeling with at least two lens mass-distance relations. Often, a constraint on the Einstein ring radius 
measurement is obtained from the caustic crossing time: This is supplemented by
secondary constraints such as precise parallax measurements and/or measures of the lens luminosity using high angular resolution observations. 
In the discovery paper of the planetary system OGLE-2014-BLG-0124Lb, the OGLE ground-based observations were combined with simultaneous Spitzer observations,
providing good measurements of the mass ratio and projeced separation of the planetary system. The parallax was also well-measured, but the photometric data
failed to tightly constrain the Einstein ring radius, $\Theta_E$.
As a consequence, the physical parameters are therefore poorly 
constrained in the original study.
We resolved the source+lens star from sub-arcsecond blends in H band using adaptive optics (AO) observations with NIRC2 mounted on Keck II telescope. 
We identify additional flux, coincident with the source to within 160 mas. We estimate the potential contributions to this blended light (chance-aligned star, 
additional companion to the lens or to the source) and find that 85\% of of the NIR flux is due to the lens star at  $\rm H_L=16.63 \pm 0.06$ 
and $\rm K_L=16.46 \pm 0.06$.  
We combined the parallax constraint and the AO constraint to derive the physical parameters of the system.
The lensing system is composed of a mid-late type G main sequence star of $\rm  M_L=0.89 \pm 0.05~M_\odot$ located 
at $\rm D_L = 3.6 \pm 0.3~$kpc in the Galactic disk.  Taking 
the mass ratio and projected separation from the original study
leads to a planet of  $\rm M_p=  0.64 \pm  0.044~M_{Jupiter}$  at $3.48 \pm 0.22~$AU. 
Excellent parallax measurement from simultaneous ground-space observations have been obtained on the microlensing event OGLE-2014-BLG-0124, but
it is only when they are combined with $\sim 30 \ $min of Keck II AO observations that the physical parameters of the host star are well measured.
\end{abstract}


\section{Mass-distance relations for microlensing}

Gravitational microlensing is unique in its sensitivity to exoplanets down to Earth mass  beyond the snow line  \citep{1991ApJ...374L..37M,1992ApJ...396..104G}, where the core accretion theory predicts that the most massive planets will form. 
However, the major limitation of most of the 51 exoplanetary microlensing analyses published to date has 
been the relatively low precision measurements of physical parameters of the system, owing to uncertain the host star mass and its distance.
By contrast, the relative physical parameters (mass ratio, projected separation relative to the angular Einstein ring radius)
are usually known with high precision. 
In the vast majority of microlensing events, the Einstein ring radius
crossing time $t_{\rm E}$ is the only measurable parameter constraining the lens mass, lens distance, and relative lens-source proper motion 
$\mu_{\rm rel}$, which are therefore degenerate. For binary microlensing events, it is possible to accurately measure the mass ratio $q$ and the
projected separation $d$ in units of Einstein ring radius. The source star often transits the caustic, providing the source radius crossing time $\rm t_*$. 
Morever, the angular radius of the source star $\theta_*$ can be estimated from the surface 
brightness relation \citep{2004AA...428..587K,Boyajian:2013kh,2014ApJ...787...92B},
so the measurement of  $\rm t_*$ yields the angular Einstein radius,  $\rm  \Theta_{\rm E} = \theta_*~t_E / t_*$


These constraints lead to a mass-distance relation betweeen lens mass $M_L$ at distance $D_L$,
with the form 
\begin{equation} \label{eq:thetaPirel}
\rm M_L = \theta_E^2 / (\kappa~\pi_{\pm rel})
\end{equation}
 
where $\rm \pi_{\pm rel} = (AU) (D_S-D_L) / (D_S~D_L)$ and $\rm \kappa=8.144~mas~M_\odot^{-1}$.

There is also a relation between the parallax $\pi_{E}$ and the mass,

\begin{equation}\label{eq:thetaPiE}
\rm M_L = \theta_E / (\kappa~\pi_{E}). 
\end{equation}

This allows the elimination of $\theta_{\rm E}$ to give a useful mass-distance relation for the case when we have well-defined parallax $\pi_{\rm E}$
but unknown  $\theta_{\rm E}$:

\begin{equation}\label{eq:nothetaE}
 \rm M_L = \pi_ {\rm rel}/ (\kappa~\pi_{E}^2)
\end{equation}
 
An independent mass-distance relation can be applied if the flux from the lens system can be reliably measured and compared to stellar models.
Using high angular resolution observations with Keck II, SUBARU or HST it is possible to separate
the contributions of the source and lens stars from blended stars at the subarcsecond level. 
We can then measure the lens apparent magnitude $ \rm m_{L}(\lambda)$ and combine
it with isochrones \citep[e.g.,][]{2008AA...484..815B} to get another mass-distance relation:

\begin{multline}\label{eq:modulus}
\rm m_L(\lambda) = 10 + 5\log(\rm D_L/1~kpc) + A_L(\lambda) \\
 + M_{\rm isochrone}(\lambda, \rm M_L, age, [Fe/H])
 \end{multline}

 
 where $M_{\rm isochrone}$ is the predicted absolute magnitude of the lens assuming 
 a given mass, age, and metallicity, and A$_L(\lambda)$ is the wavelength-dependent
 interstellar extinction along the line of sight to the lens.
 
 In practice, the parallax vector is often not well constrained and there is a degeneracy with orbital motion,
 while the Einstein ring radius is usually known to about 10\%. Therefore,
 it is quite common to combine the mass-distance relations from adaptive optics (Eq.~\ref{eq:modulus}) and $\theta_{\rm E}$ (Eq.~\ref{eq:thetaPiE}) to  measure the masses. This has been done on a number of planetary microlensing events \citep{2010ApJ...711..731J,2012A&A...540A..78K,2014ApJ...780...54B, 2015ApJ...808..170B, 2015ApJ...808..169B}.
In the favorable cases, it is possible to constrain the physical parameters of the system to within $\approx$5\%.
Recently, \citet{2016arXiv160703267K} presented the discovery of a sub Saturn-mass planet and estimated the 
mass by combining parallax measurements (Eq.~\ref{eq:nothetaE}) and adaptive optics measurement, 
without a good measurement of $\theta_{\rm E}$. In this particular case, the accuracy of the parallax is the limiting factor determining
the accuracy of the derived physical parameters. This is often the case for ground-based measurements, where only the 
parallax component parallel to the Earth acceleration is well-measured, while the other is uncertain. 

In order to overcome this limitation, a natural way forward is to obtain accurate parallaxes, by making simultaneous ground and space
observations, as proposed first by \citet{1966MNRAS.134..315R} and further developed by \citet{1992ApJ...392..442G}.  
In contrast with observations from the ground alone, both components of the parallax vector could be well-constrained. 
Three observing campaigns of simultaneous ground-based and {\sl Spitzer}  observations were completed in 2014-2016 \citep{2015ApJ...810..155Y,2015ApJ...802...76Y,2015ApJ...814...92C}.
 
 \section{The planetary system OGLE-2014-BLG-0124}

A very favorable case was OGLE-2014-BLG-0124, in which a system with a star plus 
planetary companion with mass ratio $q\sim 7\times10^{-4}$  and projected
separation $d  \sim 0.94$ was detected by the OGLE survey. It was observed simultaneously by a fleet  of ground-based telescopes 
(MOA, LCOGT, Wise 1m, MINDSTEP and SAAO 1m), and 
by the {\sl Spitzer} space telescope. It should be emphasized that this event was very favorable 
for parallax detection given its long time scale. 
During the ongoing microlensing event, models were circulated to characterize the nature of the event
and optimize requests for complementary observations from follow-up telescopes.

\begin{figure}
\vspace{5mm}
\includegraphics[scale=1,angle=0]{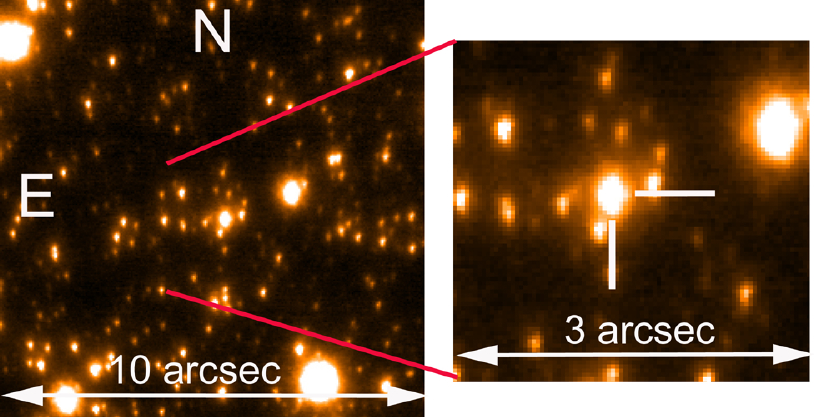}
\caption{ Keck II H-band observation of  OGLE-2014-BLG-0124. At the position of the source, we detect significant 
additional flux within 150 mas which is most likely the lens.
The elongation seen on the image is consistent with the PSF shape of other nearby field stars. \label{fig1}}
\end{figure}

\citet{2015ApJ...799..237U} presented the analysis of OGLE and {\sl Spitzer} data. OGLE captured 
the overall geometry of the microlensing  light curve, a source transiting close to a resonant caustic.
Although their model was ultimately based on OGLE and {\sl  Spitzer} data alone, it was completely
consistent with the models created during the event including data collected by the fleet of follow-up telescopes.
Unfortunately, the original study failed to acknowledge this contribution from the community.

The OGLE data on its own allowed a $\pi_{\rm E}$ measurement to $\pm$20\%.  
The inclusion of {\it Spitzer} data improved this by a factor 7, making OGLE-2014-BLG-0124
the most precise microlensing parallax measurement to date. 
Unfortunately, the trajectory of the source star did not make any caustic crossings; \citet{2015ApJ...799..237U} showed that ${\rm t_*}$ is uncertain, 
which transfers into a poorly known $\theta_{\rm E}$, and hence a large uncertainty on physical mass of the host star. 
It was also discussed by \citet{2015ApJ...814L..11Y}.

As a result, the system has two published solutions, which overlap within the errorbars. The first,
for $\rm u_0>0$,  has an $\rm M = 0.71 \pm 0.22~M_\odot$ star located at $\rm 4.1 \pm 0.6~$kpc in the galactic disk, orbited by a planet of $\rm M=  0.51 \pm  0.16~M_{Jupiter}$  at $3.1 \pm 0.5~$AU; The second one $\rm (u_0<0)$, has an $\rm M = 0.65 \pm 0.22~M_\odot$ star located at $\rm 4.23 \pm 0.59~$kpc in the galactic disk, 
orbited by a planet of $\rm M=  0.47 \pm  0.15~M_{Jupiter}$  at $2.97 \pm 0.51~$AU. We remark that the microlensing parameters (mass-ratio, projected
separation) are very close, and that the small difference in the physical parameters is coming mostly from the Bayesian modelling.

\subsection{Source star properties}

 \citet{2015ApJ...799..237U}  fitted the source magnitude $ \rm I_S = 18.59  \pm 0.02$ with a bright blend contribution of $ \rm I_{Blend} = 17.79 \pm 0.01$. 
They estimated  the extinction to be $ \rm A_I=1.02$, so $ \rm A_H=0.236$ and  $ \rm A_K=0.17$, which leads to a dereddened source color of $ \rm (V-I)_{S0}  =0.70 $. 
Using the relations in \citet{1988PASP..100.1134B} we derived  $\rm (I-H)_{S0}  =0.765$ and $\rm (H-K)_{S0}  =0.055$. 
Knowing the extinction in the different bands, we predict the source magnitudes to be $\rm H_S = 17.04  \pm 0.05$ and $\rm K_S = 16.985  \pm 0.05$.

A direct measurement of the near-infrared magnitude of the source$+$lens therefore allows us to find the flux of the lens, and
then to use Equation~\ref{eq:modulus} to get a new mass-distance relation.
 We follow \citet{2015ApJ...808..169B} and \citet{2016ApJ...824...83B} to estimate the extinction towards the lens. 
 We adopt as a scale height of the dust  towards the galactic bulge  $\rm \tau_{dust} = (0.120 kpc)/sin(b)$, where $b=-2.9167^o$ is the galactic latitude. 
Then we write the lens extinction A$_{\rm H_L}$:
\begin{equation}
\rm A_{H_L}=(1-e^{ -{  D_L/ \tau_{dust}}  }) / (1-e^{ -{  D_S/ \tau_{dust}}  })  A_{H_S}.
\end{equation}

\subsection{VVV K band light curve of OGLE-2014-BLG-0124}
We extracted  $\rm H$ and $\rm K$ cubes of images centered of the target collected by the 4m VISTA telescope at Paranal during the 
VVV survey \citep{2010NewA...15..433M}. The data set is composed of 1 H and 312 K band epoch. Using our standard procedure 
we perform PSF photometry on all the frames and calibrated them \citep{2016ApJ...824...83B,2017Marquette}.  
We fitted OGLE and VVV data using a Markov Chain Monte Carlo in order to derive an estimate of the K band calibrated source flux.
We derived $\rm K_S(fit) = 17.007 \pm 0.038$, blended with  $15.964  \pm 0.014$. This is very close and in agreement with our 
estimates from previous section of $\rm K_S = 16.985  \pm 0.05$.

Since it is a direct measurement, we adopt in the following  the fitted K band source flux to be 
$\rm K_S = 17.007 \pm 0.038$. The difference between the direct measurement and the extrapolation
reflects the level of systematics errors in our procedure. We keep the H band estimate with $\rm H_S = 17.04  \pm 0.05$.

\subsection{Keck II adaptive optics observations of OGLE-2014-BLG-0124}

On August 4, 2016 we observed OGLE-2014-BLG-0124 using NIRC2 mounted on the Keck II telescope on Mauna Kea.
We used the wide camera, with a pixel scale of 0.04 arcsec.
 We took 2 frames with an exposure time of $3 \times 10~$ sec at each of the 5 dithered positions in $\rm H$ and $\rm K$.
We followed the data reduction and calibration procedures described by \citet{2016ApJ...824...83B}.

We correct for dark and flatfield using standard procedures and stack the images
using SWarp from the Astr{\sl O}matic suite of astronomy tools \citep{2010ascl.soft10068B}.  
We cross identify the VVV and the Keck II sources and estimate the calibration constant. We estimate the uncertainty
on the zeropoint to be $0.008~$mag in $\rm H$ and $0.01$ in $\rm K$. We apply this zeropoint to the Keck II catalogues.

We identify the source+lens  star at the position marked on Figure~ \ref{fig1}. It has several blends at the $\sim 2$ arcsec 
level. The total magnitude is $H_{VVV} = 15.75 \pm 0.07$ and $K_{VVV} = 15.66 \pm 0.10$ in the VVV images. 

At the predicted position of the source, we measure  $H_{\rm Keck} = 15.95 \pm 0.04$ and  $K_{\rm Keck} = 15.79 \pm 0.03$. The PSF is slightly elongated due to the observing conditions; the ellipticity is identical to  the PSF of nearby stars. Since the source has $\rm H_S = 17.04  \pm 0.05$ and $\rm K_S = 17.007  \pm 0.038$, we estimate the blended light to be $\rm H_{\rm Blend} = 16.45  \pm 0.06$ and $K_{\rm Blend} = 16.22 \pm 0.04$. 


\section{Lens star properties}

We detected blended light aligned with the source to the order better than the 160 milliarcsecond PSF full-width at half-maximum, 
so we must estimate if it is likely to be the lens star alone, or has the contributions from:\\
- The lens \\
- An ambient star (aligned with source and lens not associated with either) \\
- A companion to the lens \\
- A companion to the source.\\
We decided to compute the contribution to the blended light using two different methods and compare them.

\subsection{Estimating contaminants, Batista et al.'s approach}

We  follow the Bayesian analysis described in \citet{2017Batista}.
First, we calculate the probability for an unrelated star in the magnitude range $\rm H=15-21$ to be aligned by chance with the lens 
and the source. We assume that stars with a separation larger than  $0.8 \times \rm FWHM Keck=$ 130 mas  would be resolved. The probabilty 
of a field star contribution to the extra NIR flux is then equal to the surface number density of stars multiplied by the area ratio 
between a circle of 130 mas and the entire field. 

While the upper limit to the separation is observationally given by 0.8$\times$FWHM, the appropriate lower limits differ between 
a lens companion and a source companion. For a lens companion, 
we consider lower limits given the absence of signature from a source and a lens companion in the light curve, 
following the approach of  \citet{2014ApJ...780...54B}. We take a conservative lower limit to the separation by considering 
the upper limit on the microlensing shear that would be induced by an additional caustic,

$$\gamma = {\rm q \over s^2 } < 10^{-3} $$

For a source companion, the lower limit is given by the minimum separation for which the companion would not produce an additional perturbation in the light curve,

$$\rm s  \ge 1/4~\theta_E~\sim~0.23~mas$$

\begin{figure}
\vspace{5mm}
\includegraphics[scale=0.425,angle=0]{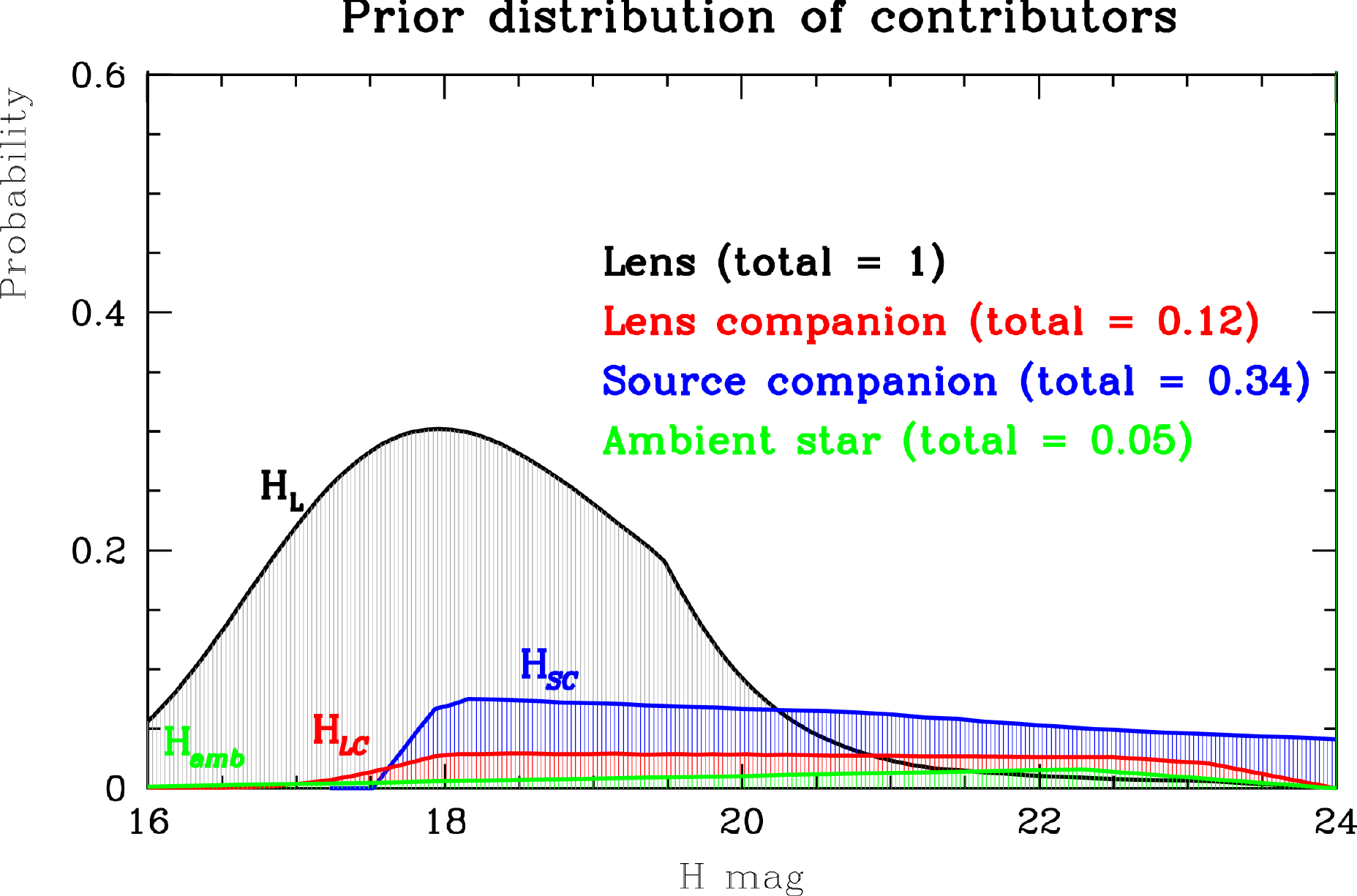}
\caption{Prior distribution of contributors in H band flux, lens, ambiant star, companion to source and companion to lens.\label{fig2}}
\end{figure}

The source and lens companions prior distributions of flux are calculated following the properties
of binary star populations described by \citet{2013ARA&A..51..269D} (see Batista et al.\ 2017 for details). The distributions of the four potential contributors (lens, ambient star, source or lens companion) are shown in Figure~\ref{fig2}.

\begin{figure}
\vspace{5mm}
\includegraphics[scale=0.75,angle=0]{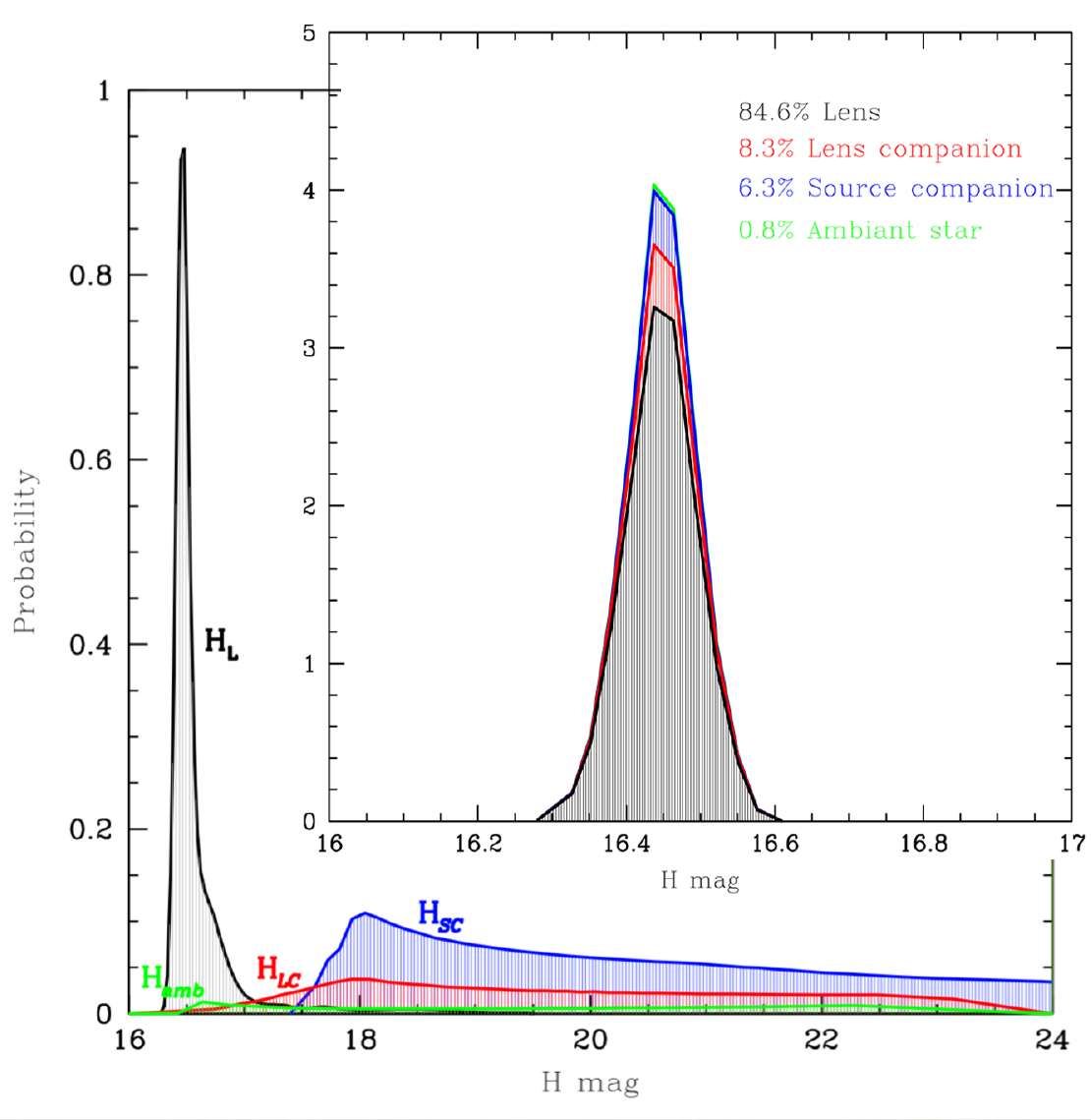}
\caption{Posterior magnitude distribution of contributors to H band flux: lens, ambient star, source companion, and lens companion. The insert 
shows the fraction of different flux sources accounting for the measured blended light. The dominant source is the lens, 
but companions to the source and the lens each have a significant expected contribution. Here 
they account for 15\% of the measured blended flux.\label{fig3}}
\end{figure}

We combine the expected flux contributions from the four potentially luminous objects into 500,000 chains, weighted by their distributions and the Keck measurement. We extract a sample
of the 1000 best fits and conclude that the most likely value of the lens contribution to the extra NIR flux is 85\%. Figure~\ref{fig3} gives 
the posterior probability distribution of the sources of extra flux, with the inset showing the most probable contribution of each source
to the detected object within a 160 mas separation. 

\subsection{Estimating contaminants, Koshimoto et al.'s approach}
The same calculation has also been done following the approach of \citet{2017AJ....154....3K}; these authors also use the multiplicity estimates from \citet{2013ARA&A..51..269D},
but the treatment of the surface density distribution of field stars is slightly different. 
The two approaches also slightly differ in their a priori distributions: Koshimoto et al.
(2017b) use a continuous law which is function of the primary mass, whereas Batista et al. (2017) use a set of distinct laws associated to different mass bins.

Moreover,  \citet{2017Naoki} use a Galactic model in their calculation, while  \citet{2017Batista} use the best fit 
parameters from  \citet{2015ApJ...799..237U} for $ \rm M,~D_L,~\Pi_E$, and $\theta_E$, and the OGLE calculator for $ \rm D_S$. 
Finally, the treatment of the Keck measurement in their Bayesian analysis slightly differs, since  \citet{2017Naoki} use it as a selection criteria of their flux combinations, while   \citet{2017Batista} use it as an a priori distribution.

Nevertheless, prior and posterior distributions are very similar and the fraction of the blended flux attributed 
to the lens is in agreement with the approach we adopted. The different contributions are estimated to be 
79.3 \% for the lens, 2.4\% for a chance aligned star, 10 \% for a companion to the source and 8 \% for a companion to the lens.
This would lead a lens less massive by $\rm \sim 0.005\ M_\odot$ than using the approach adopted in the previous paragraph.
We repeat the same calculation for the K band data, and obtain very similar results. 

We conclude that the lens contributes to the great majority of the excess NIR flux detected in the Keck adaptative optics images, regardless of minor variations to the  calculation of contamination probabilities.



\section{Discussion and conclusions}

\begin{figure}
\vspace{5mm}
\includegraphics[scale=1.,angle=0]{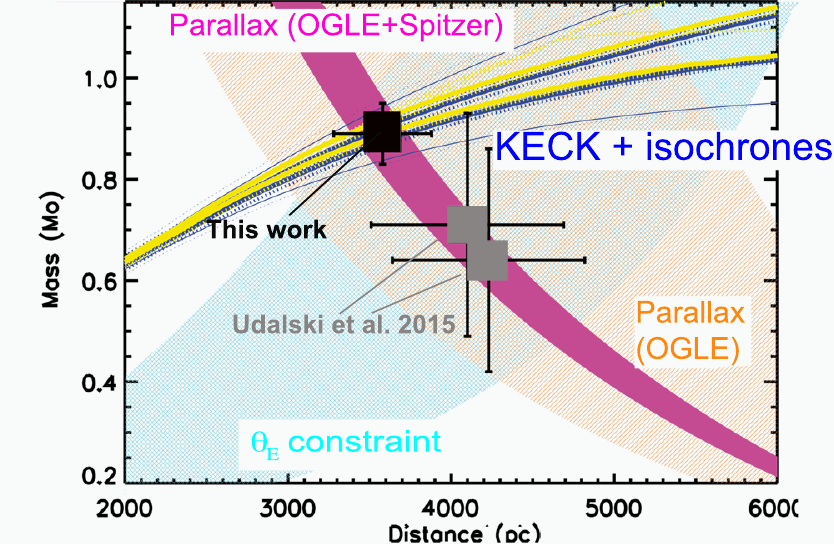}
\caption{ H-band isochrones in blue \citep{2008AA...484..815B} for the $H_L = 16.63 \pm 0.06$ lens brightness
for first planetary event with a Spitzer microlensing parallax measurement, OGLE-2014-BLG-0124. K-band isochrones
for $H_K = 16.46 \pm 0.06$ are plotted in yellow.
The parallax mass-distance relation from OGLE alone and OGLE + Spitzer are shown in orange and magenta respectively.
The mass-distance from Einstein ring radius $\rm \theta_E$ estimate is shown in cyan.
 The grey squares marks the mass and distance estimates for the 2 solutions presented in the discovery paper.
 We plot our estimate as a black square with its error bar. \label{fig4}}\vskip 1.0 truecm
\end{figure}

We estimated that 85\% of the blended flux is due to the lens in H, therefore $H_L=16.63 \pm 0.06$. Similar result 
is obtained for K, so $K_L = 16.46 \pm 0.06$.
We present in  Figure \ref{fig4} the constraints on mass and distance obtained for OGLE-2014-BLG-0124, via the 3
different routes summarized in equations (2, 3, 4), namely parallax, constraint on $\rm \theta_E$ and 
measuring the light from the lens.
First, we use the mass-distance relation from  $\rm \theta_E$ and OGLE parallax; this gives a poor constraint on mass and 
distance of the system. However, the parallax constraint coming from OGLE combined with {\it Spitzer} is much stronger as 
drawn in pink. The grey squares indicate the two solutions for $u_0 > 0$ and $u_0 <0$ presented 
by \citep{2015ApJ...799..237U}, combining the accurate ground-space parallax with 
the mass-distance relation from $\rm \theta_E$ (blue band). The latter constraint is quite weak
due to the absence of caustic crossings in the source trajectory, with the consequent uncertain fitted value of ${\rm t}_*$.
 
Our solution, plotted as a black square, relies  on well determined parameters from adaptive optics measurements and {\it Spitzer} parallax, and is in good agreement with the loose $ \theta_{\rm E}~$ constraint. The lens star is a $\rm M_L = 0.89 \pm 0.05~M_\odot$  at a distance of  $\rm D_L = 3.5 \pm 0.2~kpc$.

At this mass, the lens star would be a typical mid-late type main sequence star in the disk. Age contraints are weak, but most compatible with a typical age for disk stars, in the range $\rm \sim 4-7~Gyr$, assuming solar metallicity.
Using the lens mass $\rm M_L$, distance $D_L$, and the parallax $\Pi_E$ we can recalculate that $\rm \Theta_{E}(calc) = 1.03 \pm $0.06 mas, corresponding to $\rm 3.69 \pm 0.21 AU$. We then use 
mass ratios and projected separation presented by \citet{2015ApJ...799..237U}  for the  $u_0 >0$ and $u_0<0$ cases. 
The two solutions for the physical parameters are very close (mutually consistent within errorbars), so we conclude that $\rm M_p=  0.64 \pm  0.044~M_{Jupiter}$  at $3.48 \pm 0.22~$AU. 

This study shows the power of high angular resolution observations for constraining the host star properties in planetary microlensing events. 
It is also a cautionary tale showing that it is important to carefully estimate the potential contribution of source and lens companions, which may
potentially bias the inferred host properties
if they are not accounted for. We note that for fainter lenses, these contributions will be more dramatic, like 
the case of MOA-2016-BLG-227  \citep{2017AJ....154....3K}; a dedicated study will have to be performed in the
framework of Euclid and WFIRST. Not accounting for these potential companions might lead to a bias towards higher inferred lens masses.
In this case, because the lens star is bright, doing so would have resulted in a host mass $\rm \sim 0.02M_\odot $larger, or $(1/2.5)\sigma$. This will become even more important in the case of fainter source and lens stars.

\bibliographystyle{apj}
\bibliography{ms}

\acknowledgments
This work was supported by the University of Tasmania through the UTAS Foundation and the endowed Warren Chair in Astronomy.
D.P.B. was supported by grants NASA-NNX12AF54G, JPL-RSA 1453175 and NSF AST-1211875. V.B. was supported by CNES.
This work was partially supported by a NASA Keck PI Data Award, administered by the NASA Exoplanet Science
Institute. Data presented herein were obtained at the W. M. Keck Observatory from telescope time
allocated to the National Aeronautics and Space Administration through the agencys scientic
partnership with the California Institute of Technology and the University of California. The
Observatory was made possible by the generous financial support of the W. M. Keck Foundation.
Work by YS and CH was supported by an appointment to the NASA Postdoctoral Program at the Jet Propulsion Laboratory, California Institute of Technology, administered by Universities Space Research Association through a contract with NASA.
This publication makes use of data products from the Two Micron All Sky Survey, which is a joint project of the University of Massachusetts and the Infrared Processing and Analysis Center/California Institute of Technology, funded by the National Aeronautics and Space Administration and the National Science Foundation. Work by N.K. is supported by JSPS KAKENHI Grant Number JP15J01676.
Work of DDP and KL was supported by the University of Rijeka Grant Number 13.12.1.3.02.



\clearpage

\end{document}